\documentclass[aps,pre,reprint,amsmath,amssymb]{revtex4-2}
\usepackage[T1]{fontenc}
\usepackage{lmodern,microtype,amsmath,amssymb,amsthm,graphicx,etoolbox}
\usepackage[hidelinks]{hyperref}
\hypersetup{pdftitle={Optimal Trading Filters and the Value of Predictive Signals},pdfauthor={Matteo Smerlak}}
\newcommand{\E}{\mathbb E}
\newcommand{\R}{\mathbb R}
\newcommand{\F}{\mathcal F}
\newtheorem{proposition}{Proposition}
\begin{document}
\title{Optimal Trading Filters and the Value of Predictive Signals}
\author{Matteo Smerlak}
\email{matteo.smerlak@cfm.com}
\affiliation{Capital Fund Management, 23 rue de l'Universit\'e, 75007 Paris, France}
\begin{abstract}
The value of a forecast depends on its accuracy, but also on its actionability. In trading, an investor must translate return forecasts into positions while accounting for trading costs and the lingering impact of earlier trades. We solve the stationary quadratic trading problem with transient price impact using tools from prediction theory: Wiener--Hopf factorization gives the optimal policy as the composition of a conditional forecast and an explicit causal linear filter. We recover the G\^arleanu--Pedersen aim-portfolio under exponential resilience and derive fractional trading filters under power-law impact decay, requiring memory beyond finite-dimensional LQR control. At fixed positive risk penalty, signals of equal instantaneous predictive strength have equal frictionless Markowitz value. With trading costs, net attainable value is determined by the mean square of the friction-weighted conditional forecast. Under power-law impact with exposure bleeding, we find an intermediate optimal mean-reversion rate for Ornstein--Uhlenbeck signals.
\end{abstract}
\maketitle
\raggedbottom

\section{Introduction}

Given predictions of future returns, an investor must determine how to act on them and what they are worth after trading costs. Transient price impact connects these decisions across time: a trade earns exposure to the predicted return but also changes the cost of subsequent trades. The investor must choose how quickly to build, maintain, and unwind a position as both forecasts and impact evolve.

Transient impact has both theoretical and empirical motivation. In the resilient-order-book model of \citet{ObizhaevaWang2013}, depleted liquidity recovers between orders. \citet{BouchaudGefenPottersWyart2004} documented persistent order signs alongside approximately diffusive prices; within their propagator model, individual-trade impact must decay with age. \citet{BouchaudKockelkorenPotters2006} inferred algebraic decay after accounting for order-flow correlations, which confound the measured response to a trade with subsequent trading. We use exponential and power-law kernels as tractable models, retaining linearity in trading rate as an assumption; the detailed decay law remains market- and estimation-dependent.

In this paper we solve the stationary quadratic version of the trading problem using tools from prediction theory. We consider ongoing investment with an established trading history and no terminal liquidation date, allowing exposure to bleed between trades. (Such exposure bleeding occurs, for example, when an options book loses vega as maturity approaches, so maintaining a fixed exposure requires costly replenishment \citep{OICVega}.) Using Wiener--Hopf factorization, we decompose the optimal policy into a conditional forecast of a friction-adjusted signal followed by a causal linear filter \citep{Wiener1949,Whittle1963,Staffans1995,SubbaRaoYang2022}. From this construction we obtain both the policy and its net value, together with the loss relative to knowing the entire predictor path. In the Gaussian examples, we obtain a linear policy with an explicit frequency response.

We recover the target portfolio of \citet{GarleanuPedersen2013,GarleanuPedersen2016} and derive fractional trading filters under power-law impact. Exponential impact admits finite-dimensional relaxation dynamics. Power-law impact gives a nonrational response, precluding an exact finite-dimensional linear time-invariant realization even for an Ornstein--Uhlenbeck (OU) predictor. These fractional policies apply to stationary investment, complementing the finite-horizon liquidation results of \citet{FordeSanchezSmith2022}. Mat\'ern predictors allow us to study history-dependent forecasting alongside market memory.

We compare signals at fixed instantaneous predictive strength. Under frictionless Markowitz allocation with a fixed positive inventory penalty \citep{Markowitz1952}, optimal value depends only on predictor variance, so all signals in this comparison have the same frictionless value. We show that attainable value equals half the mean square of a friction-weighted conditional forecast of future drift. This gives a market-dependent criterion for comparing predictors: their value depends on how strongly they forecast the future drift components that the investor can profitably exploit.

Under power-law impact, replenishing bleeding exposure penalizes persistent positions, while adjustment costs penalize rapid changes. The balance selects an intermediate optimal mean-reversion rate for OU signals. Coherence provides a separate source of value: concentrating signal variance at a frequency of minimum friction gives a sharp bound over stationary signal laws. Perfectly predictable oscillations at costly frequencies can nevertheless have arbitrarily small value. We also compare predictor smoothness after optimizing signal speed, distinguishing this comparison from holding the predictor decay rate fixed. Empirical use requires estimating predictive strength jointly with signal dynamics and trading costs.

The quadratic trading problem extends the single-period mean--variance framework of \citet{Markowitz1952} and dynamic portfolio theory \citep{Mossin1968,Samuelson1969,Merton1969}. Execution costs are studied by \citet{BertsimasLo1998,AlmgrenChriss2001,GatheralSchiedSlynko2012}. \citet{HenryBouchaud2007, GarleanuPedersen2013,GarleanuPedersen2016} study dynamic portfolio choice with predictable returns and trading frictions, including exponentially relaxing impact. \citet{LehalleNeuman2019} incorporate predictive signals into finite-horizon liquidation with transient impact; \citet{NeumanVoss2022} derive signal-adaptive execution policies with temporary and transient impact. \citet{AbiJaberNeuman2025} treat general Volterra kernels and progressively measurable signals, while \citet{AbiJaberNeumanTuschmann2026} study portfolio choice with transient cross-impact. Our stationary formulation gives explicit trading filters and comparisons of signal value across speeds and coherence levels.

In Section~II we derive the general solution; in Section~III we obtain explicit policies. We compare signal values in Section~\ref{sec:value} and discuss extensions in Section~\ref{sec:extensions}. Derivations, technical assumptions, and general Bessel market extensions are given in the appendices.

\section{Optimal trading filters}\label{sec:model}
In this section we define and solve the stationary quadratic trading problem. We interpret the optimal policy as a causal prediction of the anticipative position, with prediction errors weighted by trading friction.

\subsection{The quadratic trading problem}

We use upright capitals for linear operators, lowercase letters for their convolution kernels, and hats for Fourier transforms: $\widehat\varphi(\omega)=\int e^{i\omega t}\varphi(t)dt$. The predictor is $\alpha$ and the position is $\pi$.

\paragraph{Predictor.}
An investor observes a centered, stationary return predictor $\alpha_t$, such that the unimpacted price $P_t$ satisfies
\begin{equation}
dP_t=\alpha_tdt+dM_t
\end{equation}
where $M$ is a martingale with respect to the available information $(\F_t)$. With the martingale noise scale held fixed, the predictor variance $s_\alpha^2 = \E[\alpha_t^2]$ measures instantaneous forecast strength, i.e. predictivity or `alpha'; we keep it fixed throughout.

\paragraph{Bleeding.}
Trading at rate $u_t$ adds to a book's current position $\pi_t$. We allow this exposure to bleed passively at rate $\beta$:
\begin{equation}\label{eq:state}
\partial_t\pi_t=u_t-\beta\pi_t,\qquad \beta\geq0.
\end{equation}
This generalization to the usual notion of position as an integral of past trades is useful when dealing with derivative instruments with finite maturity. For example, an options portfolio can lose vega as maturity approaches, requiring additional trades to maintain its exposure \citep{OICVega}. For stock holdings the bleeding rate is $\beta = 0$.

\paragraph{Risk aversion.}
A quadratic inventory penalty $\lambda\E[\pi_t^2]/2$, with $\lambda\geq0$, expresses aversion to risk. We use inventory risk or exposure bleeding to regularize the low-frequency behavior of the examples below.

\paragraph{Impact cost.}
Trading moves prices. Let $I_t$ denote the displacement with respect to the unimpacted price $P_t$ due to present and past trading, with Volterra kernel $g$, and $C_t$ the corresponding impact cost:
\begin{equation}\label{eq:executionprice}
I_t=\int_0^\infty g(\tau)u_{t-\tau}d\tau,
\qquad C_t=I_tu_t.
\end{equation}
This is the Bouchaud \emph{et al.} propagator model of price impact \citep{BouchaudGefenPottersWyart2004,GatheralSchiedSlynko2012}. We extend the impact kernel $g(\tau)$ evenly to negative lags, with  Fourier transform $\widehat g(\omega)$. Its high-frequency limit $\eta=\lim_{|\omega|\to\infty}\widehat g(\omega)$ measures instantaneous trading costs, while finite-frequency components measure transient impact.

We assume $\widehat g\geq0$, such that for any finite trading window
\begin{equation}\label{eq:positiveimpact}
\int_{-\infty}^{\infty}
\widehat g(\omega)|\widehat u(\omega)|^2d\omega\geq0.
\end{equation}
This positivity condition rules out profits from one's own impact on round trips, the no-price-manipulation of \citet{Gatheral2010,GatheralSchiedSlynko2012}.
When $g(\tau)$ is also nonnegative, nonincreasing, and convex for $\tau>0$, \emph{transaction-triggered price manipulation} in finite-window execution is also excluded: inserting trades opposite to the order's direction cannot reduce expected impact costs \citep{AlfonsiSchiedSlynko2012,GatheralSchiedSlynko2012}.

\paragraph{Quadratic friction operator.}
The investor maximizes expected signal earnings net of inventory risk and impact costs:
\begin{equation}\label{eq:objective}
\begin{aligned}
J(\pi; \alpha)={}&\E[\alpha_t\pi_t]-\frac\lambda2\E[\pi_t^2]\\
&-\frac12\int_{\R}g(\tau)\E[u_tu_{t-\tau}]\,d\tau.
\end{aligned}
\end{equation}
The factor $1/2$ converts the causal impact integral to a symmetric one. Write $\|\pi\|_F^2$ for the total quadratic penalty, so $J(\pi, \alpha)=\E[\alpha_t\pi_t]-\|\pi\|_F^2$. The corresponding positive, self-adjoint \emph{friction operator} $\mathsf F$ is defined by $\|\pi\|_F^2=\langle\pi,\mathsf F\pi\rangle/2$, with $\langle x,y\rangle=\E[x_ty_t]$. Stationarity makes $\mathsf F$ a convolution operator, $\mathsf F\pi=f*\pi$, with \emph{friction kernel} $f$.

From the position dynamics we have $\widehat u=(\beta-i\omega)\widehat\pi$, and hence the Fourier symbol of $\mathsf F$:
\begin{equation}\label{eq:frictionsymbol}
\widehat f(\omega)=\lambda+(\beta^2+\omega^2)\widehat g(\omega).
\end{equation}
Here $\lambda$ penalizes inventory, $\beta^2\widehat g$ charges for replenishing bleeding exposure, and $\omega^2\widehat g$ charges for trade-induced price impact. We assume $\inf_\omega\widehat f(\omega)>0$ for the stationary solution; selected noncoercive limits are discussed below.

\paragraph{Adaptedness.}
A causal, or adapted, strategy uses only information available at the time of trading: $\pi_t$ must be $\F_t$-measurable. For the explicit filters and signal comparisons below, we assume that the trader forecasts future signal values using the signal's current and past values, $\{\alpha_s:s\leq t\}$, without additional predictive information. The signal itself may already incorporate prices, order flow, or other data. The general conditional-expectation solution allows a richer information set. We work on a two-sided stationary probability space with a shift-covariant filtration, so conditional projection preserves joint stationarity. For a square-integrable process $y$, define
\[
(\mathsf Py)_t=\E[y_t\mid\F_t].
\]
Conditional expectation is the orthogonal projection onto the square-integrable observables available at time $t$. Thus an adapted position satisfies $\mathsf P\pi=\pi$. Removing adaptedness while keeping the same quadratic functional $J$ gives the anticipative benchmark $\pi_{\rm ant}=\mathsf F^{-1}\alpha$. Its value is an upper bound for the adapted problem.

We optimize over the space $\mathcal A$ of adapted, jointly stationary positions with finite friction:
\begin{equation}\label{eq:problem}
v^\star(\alpha)=\sup_{\pi\in\mathcal A}J(\pi;\alpha).
\end{equation}
This optimal value is the adapted Legendre-Fenchel dual of the squared friction norm $\|\cdot\|_F^2$. The domain, rate regularity, and quadratic-form conventions for $\mathcal A$ and the operator equations below are specified in Appendix~\ref{app:filterderivations}.

\subsection{Wiener-Hopf solution}

Since $\mathsf P\pi=\pi$ for an adapted position and $\mathsf P\alpha=\alpha$, the first-order condition for \eqref{eq:problem} is
\begin{equation}\label{eq:compressedproblem}
(\mathsf P\mathsf F\mathsf P)\pi^\star=\alpha.
\end{equation}
Thus we must invert the \emph{compression} of $\mathsf F$ to adapted observables. All inverses of this compression are taken on that subspace.

Wiener--Hopf factorization makes this inversion explicit \citep{Wiener1949,Whittle1963}. Under the spectral and domain conditions specified below \citep{GohbergKrein1970}, we factor the positive friction operator as
\begin{equation}\label{eq:operatorfactor}
\mathsf F=\mathsf F_-\mathsf F_+,
\qquad \mathsf F_-=\mathsf F_+^*,
\end{equation}
where the factor $\mathsf F_+$ and its inverse are causal: they preserve adaptedness. Its adjoint $\mathsf F_-$ is anticausal and acts on future observations.

For a positive scalar friction spectrum, the logarithmic integrability condition
\begin{equation}\label{eq:szego}
\int_{-\infty}^{\infty}\frac{|\log\widehat f(\omega)|}{1+\omega^2}\,d\omega<\infty
\end{equation}
gives an outer spectral factor $\widehat f_+$ with $\widehat f=|\widehat f_+|^2$ \citep{Bingham2021}. To use this factor on stationary processes, we also require $\mathsf F_+$ to map $\mathcal A$ onto the adapted square-integrable observables, with a causal inverse \citep{Staffans1995}. For power-law impact with $\beta>0$, the observable space excludes invariant components because $\widehat f$ diverges at zero; the centered OU and Mat\'ern examples satisfy this restriction.

On the adapted subspace, the friction form factors through the restricted causal operator. For the fixed filtration, we define the optimal-response operator by
\begin{equation}\label{eq:compressedinverse}
\mathsf K:=(\mathsf P\mathsf F\mathsf P)^{-1}
=\mathsf F_+^{-1}\mathsf P\mathsf F_-^{-1}.
\end{equation}
To solve \eqref{eq:compressedproblem}, we apply the anticausal inverse to the signal, take its conditional expectation, and reconstruct the position with the causal inverse.

Under the factorization and domain conditions above, we obtain the unique optimizer and its value:
\begin{equation}\label{eq:solution}
\begin{gathered}
\pi^\star=\mathsf K\alpha,\\
v^\star=\|\pi^\star\|_F^2
=\tfrac12\E\bigl[(\mathsf P\mathsf F_-^{-1}\alpha)_t^2\bigr].
\end{gathered}
\end{equation}
Gross signal gain is twice the optimal net value:
\begin{equation}\label{eq:optimalgaincost}
\E[\alpha_t\pi_t^\star]=2\|\pi^\star\|_F^2=2v^\star.
\end{equation}
The gap to the anticipative benchmark for the same functional is
\begin{equation}\label{eq:gap}
v_{\rm ant}-v^\star=\tfrac12\E\big|(I-\mathsf{P})\mathsf F_-^{-1}\alpha\big|^2.
\end{equation}
We can express the optimization as a prediction problem. Completing the square gives
\begin{equation}\label{eq:prediction}
J(\pi;\alpha)=v_{\rm ant}-\|\pi-\mathsf F^{-1}\alpha\|_F^2,
\end{equation}
so we seek the closest adapted position to $\pi_{\rm ant}=\mathsf F^{-1}\alpha$, with errors measured in the friction norm. Factorization converts this loss into an ordinary mean-square error:
\[
\|\pi-\pi_{\rm ant}\|_F^2
=\tfrac12\E\bigl|(\mathsf F_+(\pi-\pi_{\rm ant}))_t\bigr|^2.
\]
Under the assumed range condition, $\mathsf F_+\pi$ ranges over all adapted square-integrable observables. Hence
\[
\mathsf F_+\pi^\star
=\mathsf P(\mathsf F_+\pi_{\rm ant})
=\mathsf P\mathsf F_-^{-1}\alpha.
\]
We predict the friction-transformed anticipative position and then apply $\mathsf F_+^{-1}$ to reconstruct exposure. Directly conditioning $\pi_{\rm ant}$ would minimize unweighted position error. In the Gaussian examples, the transformed prediction is linear and is a Wiener causal estimate. The conditional-expectation formulation also applies to non-Gaussian signals.

Write $h_\pm$ for the kernels of the inverse operators: $\mathsf F_\pm^{-1}x=h_\pm*x$ and $\widehat h_\pm=1/\widehat f_\pm$. The identity $\mathsf F_-^{-1}=(\mathsf F_+^{-1})^*$ gives $h_-(s)=h_+(-s)$, so $\mathsf F_-^{-1}$ weights future signals. The forecast in \eqref{eq:solution} is therefore
\begin{equation}\label{eq:conditionalforecast}
\begin{aligned}
(\mathsf F_+\pi^\star)_t&=\int_{-\infty}^0 h_-(r)\,\E[\alpha_{t-r}\mid\F_t]\,dr\\
&=\int_0^\infty h_+(s)\,\E[\alpha_{t+s}\mid\F_t]\,ds.
\end{aligned}
\end{equation}
\subsection{Transfer function and spectrum}

For a stationary predictor with spectral density $S_\alpha$ satisfying the logarithmic integrability condition \eqref{eq:szego} with $\widehat f$ replaced by $S_\alpha$, write $S_{\alpha,+}$ for its causal outer factor:
\[
S_\alpha(\omega)=|S_{\alpha,+}(\omega)|^2.
\]
This factorization depends only on second-order statistics and applies to non-Gaussian predictors as well. The factor maps unit-spectral-density white-noise innovations, understood in the linear prediction sense, to the predictor. The subscript $+$ includes the factor's phase.

We now assume the predictor is Gaussian, with innovations containing the same past information as the observed signal. Best linear prediction then coincides with conditional expectation, and the response $\mathsf K\alpha$ has a causal convolution representation $\pi^\star=k*\alpha$. The kernel $k$ depends on both friction and the predictor law; $\widehat k$ is its frequency response. Conditioning on past innovations retains nonnegative lags in their response kernel. Denoting this causal projection by $[\,\cdot\,]_+$, we obtain
\begin{equation}\label{eq:signaltransfer}
\begin{gathered}
\widehat{\pi^\star}=\widehat k\widehat\alpha,\\
\widehat k=\frac{1}{\widehat f_+ S_{\alpha,+}}
\left[\frac{S_{\alpha,+}}{\widehat f_-}\right]_+.
\end{gathered}
\end{equation}
Inside the projection we apply the anticausal inverse friction to the predictor's innovations factor. We then apply the causal inverse friction and express the response in terms of the observed predictor. All factors are evaluated at frequency $\omega$. For non-Gaussian predictors, the general solution \eqref{eq:solution} still applies, but conditional forecasts can be nonlinear and the spectrum alone need not determine the optimal adapted policy.

For the Gaussian predictor, using $\widehat f=|\widehat f_+|^2$, we obtain the position power spectral density $S_{\pi^\star}$
\[
 S_{\pi^\star}(\omega)
=\frac1{\widehat f(\omega)}
\left|\left[\frac{ S_{\alpha,+}}{\widehat f_-}\right]_+(\omega)\right|^2.
\]
The optimal value equals the friction cost of this position:
\begin{equation}\label{eq:innovations}
v^\star=\|\pi^\star\|_F^2
=\frac1{4\pi}\int_{-\infty}^{\infty}
\widehat f(\omega) S_{\pi^\star}(\omega)\,d\omega.
\end{equation}

\section{Explicit solutions}\label{sec:besseltrading}

In this section we specialize the general solution to Markov predictors, recover the G\^arleanu--Pedersen partial-adjustment and two-EMA policies, then derive fractional trading under power-law impact. We then introduce history-dependent Mat\'ern predictors and compare the two sources of memory in Fig.~\ref{fig:filters}. Derivations and cutoff extensions are collected in Appendix~\ref{app:filterderivations}.

\subsection{Markov predictors}\label{sec:oufilter}

For the Markov examples, the forecast in \eqref{eq:conditionalforecast} depends only on the current signal $\alpha_t$. For OU signals it is proportional to $\alpha_t$. Define the mean-reversion rate $\theta_s>0$ through its conditional forecast:
\begin{equation}\label{eq:ou}
(\mathsf P\alpha_{\cdot+s})_t=e^{-\theta_s s}\alpha_t,
\qquad s\geq0.
\end{equation}
For an arbitrary admissible causal operator $\mathsf F_+$ with symbol $\widehat f_+$, the optimal policy, value, and transfer are
\begin{equation}\label{eq:ouvalue}
\begin{gathered}
\pi^\star=\frac{\mathsf F_+^{-1}\alpha}{\widehat f_+(i\theta_s)},\qquad
v^\star(\theta_s)=\frac{s_\alpha^2}{2\widehat f_+(i\theta_s)^2},\\
\widehat k(\omega)=[\widehat f_+(i\theta_s)\widehat f_+(\omega)]^{-1}.
\end{gathered}
\end{equation}
Changing $\theta_s$ therefore changes only the overall gain of $\widehat k$; the normalized magnitude and phase depend only on market parameters.

\subsection{G\^arleanu--Pedersen target portfolios}\label{sec:oubenchmark}\label{sec:twoema}

\citet{GarleanuPedersen2013,GarleanuPedersen2016} derive partial adjustment toward a forward-looking target portfolio. Their continuous-time model includes both temporary costs and exponentially relaxing impact \citep[Sec.~1.2]{GarleanuPedersen2016}. For a scalar OU predictor, we express their zero-discount, zero-funding stationary specializations as a single EMA under instantaneous costs and a two-EMA policy with exponential impact, using the factorization in \eqref{eq:ouvalue}.

\paragraph{Instantaneous costs: the EMA solution.}
A familiar risk-controlled policy follows by taking ordinary holdings, $\beta=0$, local execution spectrum $\widehat g=\eta>0$, and inventory penalty $\lambda>0$. Writing $b=\sqrt{\lambda/\eta}$ gives $\widehat f=\lambda+\eta\omega^2$ and $\widehat f_+(z)=\sqrt\eta(b-iz)$. The OU position and value are
\begin{equation}\label{eq:localbenchmark}
\begin{gathered}
\partial_t\pi^\star+b\pi^\star=\frac{\alpha}{\eta(\theta_s+b)},\\
v^\star=\frac{s_\alpha^2}{2\eta(\theta_s+b)^2}.
\end{gathered}
\end{equation}
This is the stationary zero-discount limit of the partial-adjustment rule in \citet[Proposition~1]{GarleanuPedersen2016}. The target portfolio is the risk-adjusted signal $\alpha_t/\lambda$, discounted by $b/(\theta_s+b)$ for the signal's expected decay. The position relaxes toward this target at rate $b$.

\paragraph{Exponential impact: the two-EMA solution.}
Add exponential transient impact to the local execution cost: $\widehat g(\omega)=\eta+\gamma/(\theta_m^2+\omega^2)$, with $\eta,\gamma,\theta_m>0$. Retaining ordinary holdings, $\beta=0$, and inventory risk $\lambda>0$ gives
\begin{equation}\label{eq:expsymbol}
\begin{aligned}
\widehat f(\omega)&=\lambda+\eta\omega^2+
\frac{\gamma\omega^2}{\theta_m^2+\omega^2}\\
&=\eta\frac{(b_1^2+\omega^2)(b_2^2+\omega^2)}
{\theta_m^2+\omega^2}.
\end{aligned}
\end{equation}
Matching numerator coefficients determines the causal spectral factor and its rates:
\begin{equation}\label{eq:expeta}
\begin{gathered}
\widehat f_+(z)=\sqrt\eta\frac{(b_1-iz)(b_2-iz)}{\theta_m-iz},\\
b_1^2+b_2^2=\theta_m^2+(\lambda+\gamma)/\eta,\\
b_1^2b_2^2=\lambda\theta_m^2/\eta.
\end{gathered}
\end{equation}
The rates satisfy $0<b_1<\theta_m<b_2$. Substituting the factor into \eqref{eq:ouvalue} and expanding in partial fractions gives
\begin{equation}\label{eq:twoema}
\begin{aligned}
\pi_t^\star={}&\frac{\theta_s+\theta_m}{\eta(\theta_s+b_1)(\theta_s+b_2)}\\
&\times\left[
\frac{w_1}{b_1}\operatorname{EMA}_{b_1}[\alpha]_t+
\frac{w_2}{b_2}\operatorname{EMA}_{b_2}[\alpha]_t\right].
\end{aligned}
\end{equation}
The moving average and weights are
\[
\begin{gathered}
\operatorname{EMA}_b[\alpha]_t=b\int_0^\infty e^{-b\tau}\alpha_{t-\tau}d\tau,\\
w_1=\frac{\theta_m-b_1}{b_2-b_1},\qquad
w_2=\frac{b_2-\theta_m}{b_2-b_1}.
\end{gathered}
\]
Both weights are positive and sum to one. This gives the two-EMA structure of \citet[Sec.~1.2, Proposition~2]{GarleanuPedersen2016}; the corresponding characteristic rates also appear in \citet[Eqs.~(5.21)--(5.22)]{NeumanVoss2022}.

\begin{figure*}[!tp]
\centering
\includegraphics[width=\linewidth]{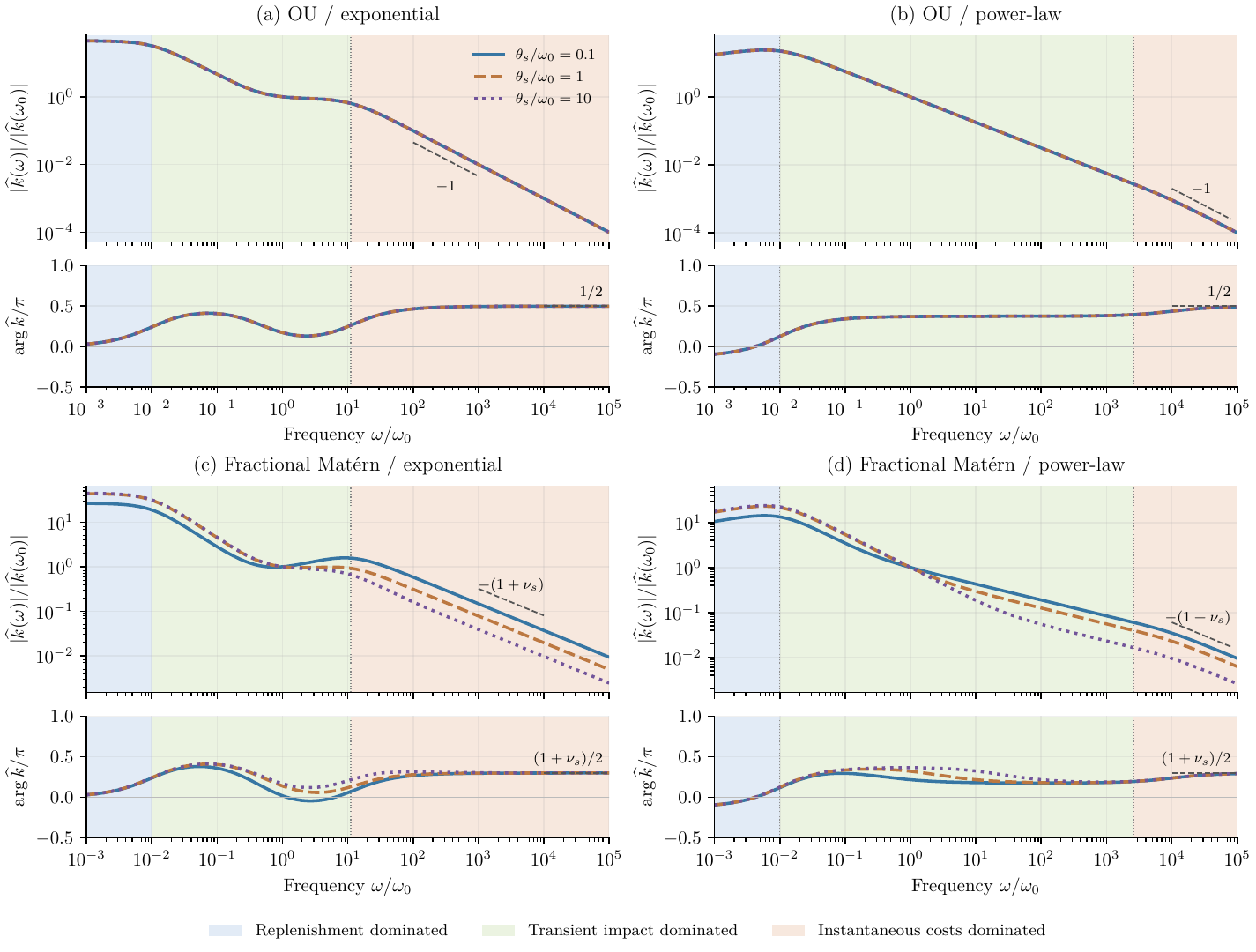}
\caption{Stationary position filters with the explicit execution-cost cutoff \eqref{eq:cutoffexecution}. Each panel shows magnitude normalized by $|\widehat k(\omega_0)|$ above and phase below. Rows use OU signals ($\nu_s=0$) and fractional Mat\'ern signals ($\nu_s=-0.4$); columns use exponential impact ($\nu_m=0$, $\theta_m/\omega_0=0.5$) and regularized power-law impact ($\nu_m=3/4$, $\theta_m=0$). Colors and line styles identify $\theta_s/\omega_0=0.1,1,10$. All panels use $\beta=\eta=0.01$, $\lambda=0$, $s_\alpha^2=1$, and $\omega_0=1$. Matching $\widehat g(\omega_0)=1$ gives $\Omega/\omega_0\simeq11.14,10^4$. OU curves coincide after normalization. Vertical bands identify the largest contribution to $\widehat f$: replenishment $\beta^2\widehat g$ (blue), transient adjustment $\omega^2(\widehat g-\eta)$ (green), or instantaneous adjustment $\eta\omega^2$ (orange). Dotted boundaries mark equality of adjacent contributions. In (b), the intermediate power-law regime has gain $|\widehat k|\propto\omega^{-\nu_m}$ and phase $\nu_m\pi/2$. Dashed guides mark the ultimate cutoff-tail magnitude exponents $-(1+\nu_s)$ and phase limits for the displayed parameters.}\label{fig:filters}
\end{figure*}

\subsection{Fractional trading under power-law impact}\label{sec:fractional}

Power-law impact yields another class of explicit solutions, with fractional trading responses. The G\^arleanu--Pedersen policies above admit finite-dimensional linear time-invariant state-space realizations; the fractional response has a nonrational transfer and therefore requires memory beyond such a representation. We derive it with the same factorization, retaining the OU predictor so that the additional memory comes entirely from impact.

Take the power-law execution spectrum $\widehat g(\omega)=\gamma|\omega|^{2\nu_m-2}$, with $\gamma>0$. The corresponding time-domain kernel is $g(t)\propto|t|^{-(2\nu_m-1)}$: its decay exponent is $2\nu_m-1$. Requiring impact to decay at long lags and remain locally integrable at zero gives $0<2\nu_m-1<1$, hence $1/2<\nu_m<1$. The broader range $0<\nu_m<1$ is available for the Bessel market with a positive low-frequency cutoff. We set $\lambda=\eta=0$ and use $\beta>0$ to obtain stationary positions.

The friction symbol is $\widehat f(\omega)=\gamma(\beta^2+\omega^2)|\omega|^{2\nu_m-2}$, with causal factor
\begin{equation}\label{eq:fracfactor}
\begin{gathered}
\widehat f_+(z)=\sqrt\gamma\frac{\beta-iz}{(-iz)^{1-\nu_m}},\\
\frac1{\widehat f_+(z)}=\frac{(-iz)^{1-\nu_m}}
{\sqrt\gamma(\beta-iz)},\qquad
\widehat f_+(i\theta_s)=\sqrt\gamma\frac{\beta+\theta_s}{\theta_s^{1-\nu_m}}.
\end{gathered}
\end{equation}
Substitution into \eqref{eq:ouvalue} gives the fractional trading rate and position:
\begin{equation}\label{eq:fracsolution}
\begin{aligned}
u_t^\star&=\frac{\theta_s^{1-\nu_m}}{\gamma(\theta_s+\beta)}\partial_t^{1-\nu_m}\alpha_t,\\
\pi_t^\star&=\frac{\theta_s^{1-\nu_m}}{\gamma(\theta_s+\beta)}
(\partial_t+\beta)^{-1}\partial_t^{1-\nu_m}\alpha_t.
\end{aligned}
\end{equation}

Thus power-law impact memory gives the optimal trading rate as a fractional derivative of the predictor: a non-local operator depending on signal values far in the past. In particular, a sufficiently declining signal can induce selling even while it remains positive, for example when reducing an existing long position. Such signal-driven adjustment is consistent with the no-manipulation conditions above, which concern gains from impact alone and fixed-quantity execution costs. The rate gain grows as $|\omega|^{1-\nu_m}$ and its phase is $-(1-\nu_m)\pi/2$ at positive frequencies. For $\beta>0$, we reconstruct a stationary position with the stable operator $(\partial_t+\beta)^{-1}$.

At $\beta=0$, integrating this rate gives no finite-variance stationary position; finite-time initialization defines a different problem.

\subsection{History-dependent predictors}\label{sec:besselkernels}\label{sec:generalfilter}\label{sec:besselfilter}\label{sec:maternexp}

History-dependent forecasts change the dynamics of the optimal position: varying signal speed can change the normalized response and its phase. We obtain a nonrational trading filter from fractional predictor memory even with local execution costs, and then examine how forecasting and market memory combine. For a stationary Gaussian Mat\'ern predictor, take $\nu_s<1/2$ and $\theta_s>0$. Its power spectrum and causal outer factor are \citep{LillySykulskiEarlyOlhede2017}
\begin{equation}\label{eq:besselpredictorspectrum}
\begin{gathered}
 S_\alpha(\omega)=a^2(\theta_s^2+\omega^2)^{\nu_s-1},\\
 S_{\alpha,+}(\omega)=a(\theta_s-i\omega)^{\nu_s-1}.
\end{gathered}
\end{equation}
The normalization $a$ fixes $\E[\alpha_t^2]=s_\alpha^2$; finite variance requires $\nu_s<1/2$, with no lower bound. At fixed exponent, $\theta_s$ sets the time scale; its correlation-time interpretation depends on $\nu_s$. Decreasing $\nu_s$ increases smoothness: the predictor has $j$ mean-square derivatives for integers $0\leq j<1/2-\nu_s$. The innovations factor is outer, so the histories of $\alpha$ and $W$ contain the same information. At $\nu_s=0$ this recovers OU; other Mat\'ern predictors have history-dependent conditional means. For Bessel impact, we give the general transfer function in \eqref{eq:besseljointtransfer}.

\paragraph{Forecasting memory with local costs.}
We first extend the local-cost benchmark \eqref{eq:localbenchmark}, with ordinary holdings $\beta=0$, $\widehat g=\eta>0$, and $\lambda>0$. Writing $b=\sqrt{\lambda/\eta}$, we obtain
\begin{equation}\label{eq:maternlocaltransfer}
\widehat k(\omega)=\frac{1-\left(\dfrac{\theta_s-i\omega}{\theta_s+b}\right)^{1-\nu_s}}
{\eta(b^2+\omega^2)}.
\end{equation}
At $\nu_s=0$, this reduces to $[\eta(\theta_s+b)(b-i\omega)]^{-1}$, the OU response. For noninteger $\nu_s$, the fractional power makes the transfer nonrational, precluding an exact finite-dimensional linear time-invariant realization. At $\nu_s=0,-1,-2,\ldots$, it is rational, but can be improper: derivatives of the smoother input are then required. Changing $\theta_s$ generally changes normalized shape and phase as well as gain, even without transient impact.

\paragraph{Combining predictor and market memory.}
For the pure power-law market of Sec.~\ref{sec:fractional}, with well-separated rates $\beta\ll\theta_s$, we obtain
\begin{equation}\label{eq:maternplscaling}
|\widehat k(\omega)|\propto
\begin{cases}
|\omega|^{-\nu_m},&\beta\ll|\omega|\ll\theta_s,\\
|\omega|^{-(\nu_m+\nu_s)},&\theta_s\ll|\omega|.
\end{cases}
\end{equation}
Above the predictor's correlation scale, its smoothness changes the response exponent; sufficiently smooth signals can have increasing high-frequency gain. These powers require separated scales and a nonzero leading forecast coefficient; without scale separation, they appear only as slope changes.

\paragraph{Comparing the responses.}\label{sec:responsecomparison}
In the upper row of Fig.~\ref{fig:filters}, the normalized OU curves coincide within each market: signal speed changes only the gain. In the lower-left panel, forecasting memory changes the shape and phase even with exponential impact. In the lower-right panel, predictor and market memory jointly determine the intermediate response, as in \eqref{eq:maternplscaling}. Positive instantaneous cost eventually replaces that scaling with the cutoff tail indicated by the dashed guides. We match the execution spectra at the reference frequency, $\widehat g(\omega_0)=1$. The plotted tail coefficients are nonzero; exceptional cancellations can change the asymptotic powers.

These responses show how forecasting memory changes the optimal use of a signal. In Section~\ref{sec:value} we compare the resulting values at fixed predictor variance.

\section{The value of predictive signals}\label{sec:value}

We hold predictor variance $s_\alpha^2=\E[\alpha_t^2]$, martingale noise scale, and market costs fixed. To interpret this normalization, consider frictionless Markowitz allocation with a fixed positive inventory penalty $\lambda_{\rm ref}$. The optimal position is $\alpha_t/\lambda_{\rm ref}$ and optimized value is $s_\alpha^2/(2\lambda_{\rm ref})$, independent of signal dynamics. Our comparison holds this auxiliary frictionless benchmark constant while varying persistence, smoothness, and coherence. Finite-horizon forecast accuracy and realized strategy Sharpe can change.

With trading costs, we evaluate each signal through its friction-weighted conditional forecast. Combining \eqref{eq:conditionalforecast} and \eqref{eq:solution}, we obtain
\begin{equation}\label{eq:signalvaluecriterion}
\begin{gathered}
z_t=\int_0^\infty h_+(s)\,\E[\alpha_{t+s}\mid\F_t]\,ds,\\
v^\star=\tfrac12\E[z_t^2].
\end{gathered}
\end{equation}
Market friction determines the weights. Among admissible signals of fixed variance, maximizing value is equivalent to maximizing this forecast's mean square under each signal's information set. Concentration at inexpensive frequencies and predictability of the relevant future drift increase value; anticipative value and causality loss distinguish these contributions. Empirical comparisons require each candidate to satisfy the conditional-drift assumption of Section~\ref{sec:model}, relative to its available information.

The figures use bleeding exposure, $\beta>0$, with $\lambda=0$. This choice matters for the conclusions. With ordinary holdings and inventory risk, $\beta=0$, $\lambda>0$, the impact models considered here have $\widehat f=\lambda+\omega^2\widehat g$, minimized at zero frequency. Slower signals are favored, and every Mat\'ern exponent approaches the same supremum $s_\alpha^2/(2\lambda)$ as $\theta_s\downarrow0$. Smoothness still orders value at a fixed predictor decay rate. Positive preferred speeds and strict differences between rate-optimized smoothness levels arise in the exposure-bleeding examples below. With $\beta=\lambda=0$, these markets have no finite bound over signal laws.

We first derive the value bound, then compare OU speeds and Mat\'ern smoothness against it. The general results use the friction and domain assumptions of Sec.~\ref{sec:model}; proofs and cutoff extensions are in Appendix~\ref{app:valuedetails}. Figure~\ref{fig:predictorpaths} shows representative signals, and Fig.~\ref{fig:valueheatmap} compares a range of speeds and exponents.

\begin{figure*}[!tp]
\centering
\includegraphics[width=\linewidth]{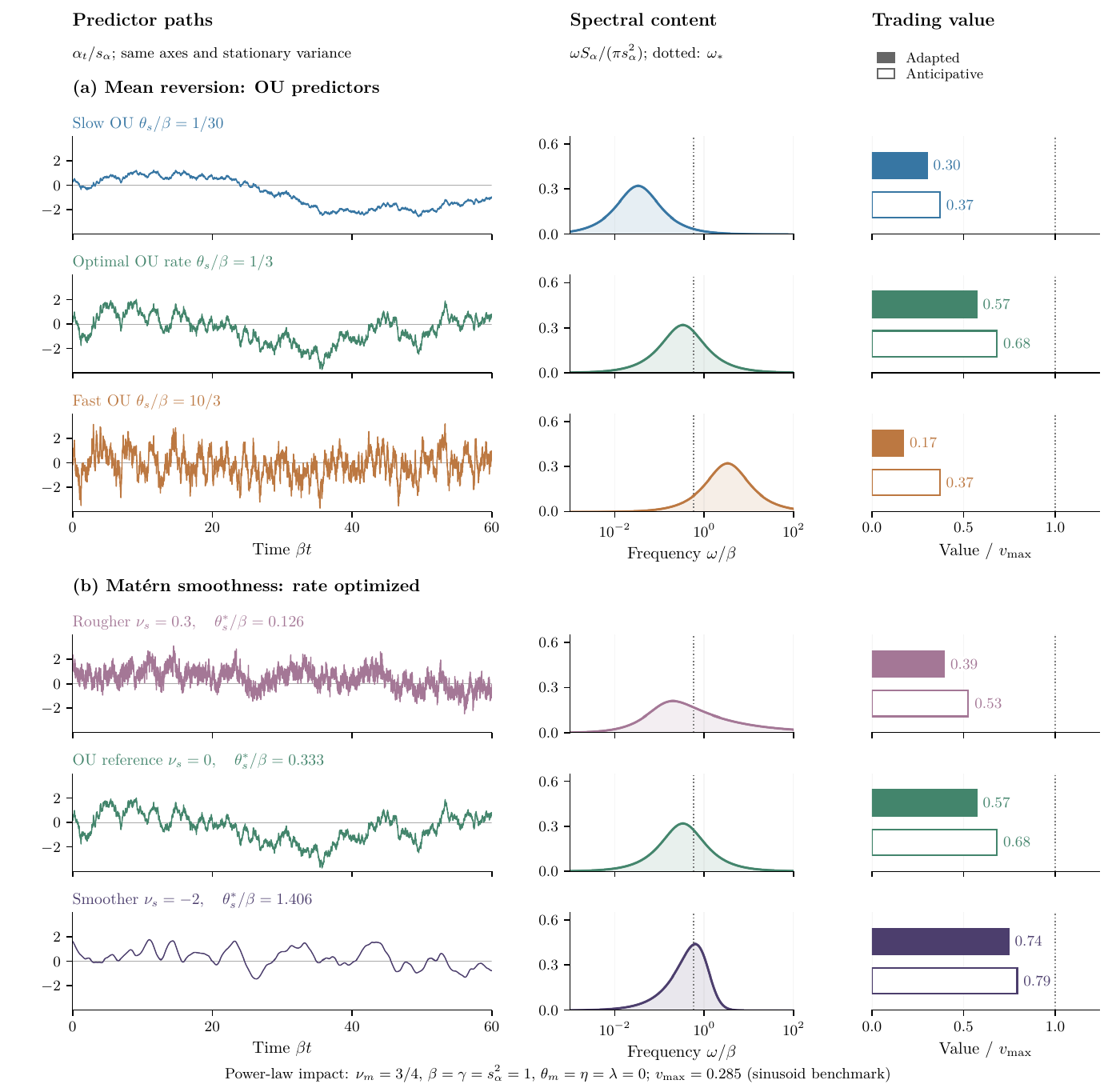}
\caption{Signal speed and Mat\'ern smoothness at fixed predictor variance and market costs. (a) OU predictors with different decay rates. (b) Mat\'ern exponents $\nu_s=0.3,0,-2$, each at its value-maximizing decay rate; the OU reference repeats the middle row of (a). Left: illustrative stationary paths on common axes. Middle: variance per logarithmic frequency interval, $\omega S_\alpha(\omega)/(\pi s_\alpha^2)$, with common axes; the dotted line marks the least-cost frequency $\omega_*$. Right: adapted (filled) and anticipative (open) stationary values divided by $v_{\max}$. The dotted value line is the separate sinusoidal bound, not a Mat\'ern endpoint. Here $\nu_m=3/4$, $\beta=\gamma=s_\alpha^2=1$, and $\theta_m=\eta=\lambda=0$. Path realizations are sampled without demeaning or sample-variance rescaling; values are computed independently of those realizations.}\label{fig:predictorpaths}
\end{figure*}

\subsection{A benchmark for signal value}\label{sec:maxsignal}

Assume $\inf_\omega\widehat f(\omega)>0$. For a stationary predictor of variance $s_\alpha^2=(2\pi)^{-1}\int_\R S_\alpha(\omega)d\omega$,
\begin{equation}\label{eq:signalbound}
\begin{aligned}
v^\star(\alpha)\leq v_{\rm ant}(\alpha)
&=\frac1{4\pi}\int_{-\infty}^{\infty}\frac{S_\alpha(\omega)}{\widehat f(\omega)}\,d\omega\\
&\leq\frac{s_\alpha^2}{2\inf_\omega\widehat f(\omega)}.
\end{aligned}
\end{equation}
The gap between the final bound and $v_{\rm ant}$ reflects the signal's spectral placement: variance at high-friction frequencies contributes less value. The gap $v_{\rm ant}-v^\star$ measures the additional loss from unavailable future observations. These are the two shortfalls shown by the open and filled bars in Fig.~\ref{fig:predictorpaths}; the paths are illustrative draws, while the bars are stationary model expectations.

A random-phase sinusoid of frequency $\omega>0$ and variance $s_\alpha^2$ is fully predictable from its past. Its value is
\begin{equation}\label{eq:predictablevalue}
v_{\rm sin}(\omega)=\frac{s_\alpha^2}{2\widehat f(\omega)}.
\end{equation}
For power-law and finite-cutoff impact, $\widehat f(\omega)\to\infty$ at high frequency, so this value can be arbitrarily small despite perfect predictability and unchanged variance.

When friction is minimized at a positive frequency $\omega_*$, we attain both inequalities in \eqref{eq:signalbound} with the stationary Gaussian sinusoid
\begin{equation}\label{eq:optimalsignal}
\begin{gathered}
\alpha_t=A\cos(\omega_*t)+B\sin(\omega_*t),\\
A,B\ \text{independent},\qquad A,B\sim N(0,s_\alpha^2).
\end{gathered}
\end{equation}
Past observations determine $A$ and $B$, making the anticipative policy feasible:
\begin{equation}\label{eq:maxsignalpolicy}
\begin{gathered}
\pi_t^\star=\frac{\alpha_t}{\widehat f(\omega_*)},\qquad
u_t^\star=\frac{\partial_t\alpha_t+\beta\alpha_t}{\widehat f(\omega_*)},\\
v_{\max}=\frac{s_\alpha^2}{2\widehat f(\omega_*)}.
\end{gathered}
\end{equation}
If the minimum occurs at zero, a centered random constant attains the bound when such observables are admissible.

For power-law impact with $\beta>0$ and $\lambda=\eta=0$, signals with square-integrable causal innovations kernels fall strictly below the bound. Damped Gaussian oscillators approach it by concentrating their spectral mass at $\pm\omega_*$. This coherence limit also holds with positive instantaneous cost.

\subsection{A preferred speed under power-law impact}\label{sec:valuespeed}

Take the power-law market of Sec.~\ref{sec:fractional}, with $1/2<\nu_m<1$, $\beta>0$, and $\lambda=\eta=0$. Its friction separates as
\[
\widehat f(\omega)=\gamma\beta^2|\omega|^{2\nu_m-2}
+\gamma|\omega|^{2\nu_m}.
\]
The first term penalizes replenishing persistent exposure; the second penalizes rapid adjustment. The OU value from \eqref{eq:ouvalue} is
\begin{equation}\label{eq:powerlawouvalue}
v^\star(0,\theta_s)=\frac{s_\alpha^2}{2\gamma}
\frac{\theta_s^{2-2\nu_m}}{(\theta_s+\beta)^2}.
\end{equation}
It vanishes for both very slow and very fast signals. Minimizing friction over carrier frequency and maximizing value over OU decay rate give
\begin{equation}\label{eq:maxsignalcomparison}
\omega_*=\beta\sqrt{\frac{1-\nu_m}{\nu_m}},
\qquad \theta_s^\star=\beta\frac{1-\nu_m}{\nu_m}.
\end{equation}
These rates solve different comparisons: $\omega_*$ selects the cheapest fully predictable oscillation, while $\theta_s^\star$ balances persistence and continual innovations within the OU family. The value bound at fixed predictor variance is
\begin{equation}\label{eq:coherentpurevalue}
v_{\max}=\frac{s_\alpha^2}{2\gamma\beta^{2\nu_m}}
\nu_m^{\nu_m}(1-\nu_m)^{1-\nu_m}.
\end{equation}
For $\nu_m=3/4$ and $\beta=\gamma=s_\alpha^2=1$, it is approximately $0.2849$, compared with $0.1624$ for the best OU predictor.

Figure~\ref{fig:predictorpaths}(a) illustrates this trade-off. The slow signal concentrates variance where replenishment is expensive; the fast signal requires costly adjustment. Their adapted values are $0.30$ and $0.17$ of the bound, versus $0.57$ at the preferred rate. Even the best OU signal retains a broad spectrum and unpredictable innovations, leaving both a spectral-placement cost and a causality loss.

With ordinary holdings and inventory risk, $\widehat f=\lambda+\gamma|\omega|^{2\nu_m}$ instead increases from a finite low-frequency penalty, favoring slow signals.

\begin{figure*}[!tp]
\centering
\includegraphics[width=\linewidth]{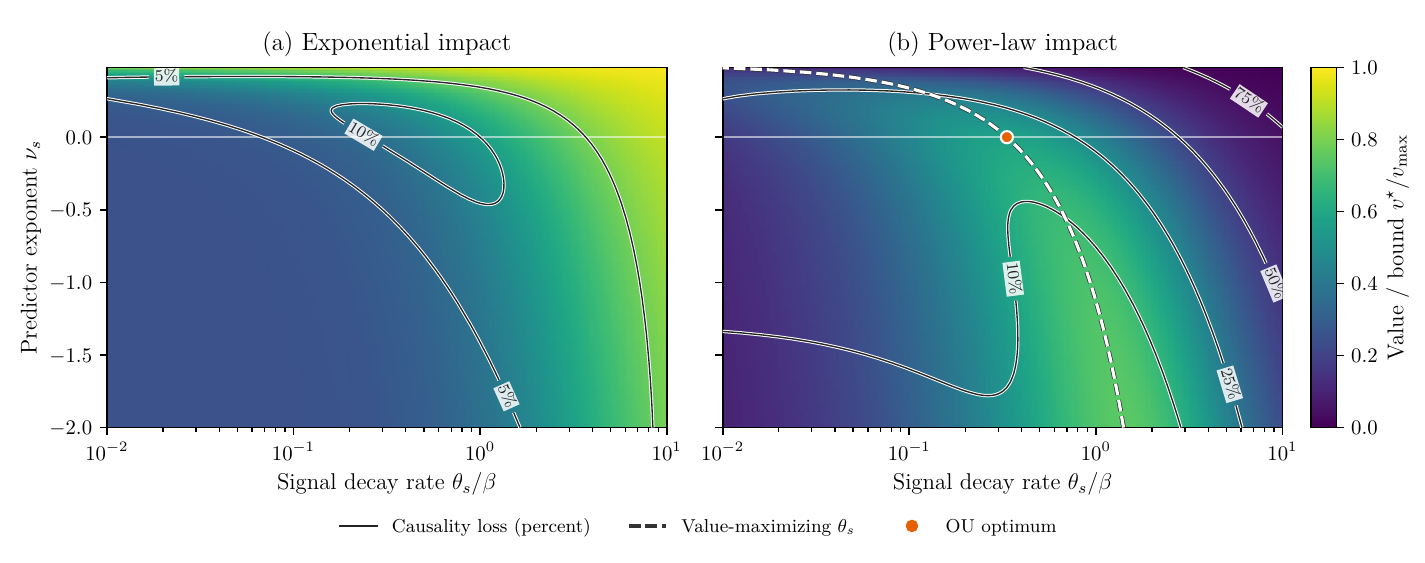}
\caption{Value and causality loss for (a) exponential and (b) power-law impact. Color shows adapted value divided by each market's bound $v_{\max}=s_\alpha^2/(2\inf\widehat f)$; thin solid contours show the fraction $(v_{\rm ant}-v^\star)/v_{\rm ant}$ lost to causality, labelled in percent. Exponential impact uses $\nu_m=0$, $\theta_m=0.5$, $\gamma=1.25$; power-law impact uses $\nu_m=3/4$, $\theta_m=0$, $\gamma=1$. Both have $\beta=s_\alpha^2=1$, $\eta=\lambda=0$, and $\widehat g(\beta)=1$. The bounds are $0.4$ and approximately $0.2849$, respectively. The horizontal line marks OU predictors. In (b), the dashed curve maximizes adapted value over the decay rate, and the orange dot marks the OU optimum. The displayed exponents extend through differentiable predictors to $\nu_s=-2$; the family continues to arbitrarily negative exponents. In (a), every exponent approaches the value bound as $\theta_s\to\infty$.}\label{fig:valueheatmap}
\end{figure*}

\subsection{The value of smoother predictors}\label{sec:preferredexponent}

For the Mat\'ern family $\nu_s<1/2$, decreasing $\nu_s$ makes the predictor smoother. We compare the best value at each exponent by optimizing its decay rate:
\begin{equation}\label{eq:exponentvalue}
\bar v^\star(\nu_s)=\sup_{\theta_s>0}v^\star(\nu_s,\theta_s).
\end{equation}
Value is nondecreasing with smoothness under this comparison. Hence the supremum over the family is
\begin{equation}\label{eq:preferredexponent}
\sup_{\nu_s<1/2,\,\theta_s>0}v^\star(\nu_s,\theta_s)
=\lim_{\nu_s\to-\infty}\bar v^\star(\nu_s).
\end{equation}
The ordering applies across admissible friction symbols; fixing the decay rate instead can reverse the ordering in the exposure-bleeding regime. For ordinary holdings with inventory risk, $\beta=0$, $\lambda>0$, the impact models considered here give a stronger comparison: smoother predictors have no lower value even at a fixed decay rate. Optimizing the rate gives the same supremum $s_\alpha^2/(2\lambda)$ for every exponent, approached by slow signals.

Figure~\ref{fig:predictorpaths}(b) shows the rougher predictor $\nu_s=0.3$, the OU reference $\nu_s=0$, and the smoother predictor $\nu_s=-2$, each at its preferred rate. They attain $0.39$, $0.57$, and $0.74$ of the value bound at fixed predictor variance. The OU row repeats the realization and value from (a). In these examples, smoother predictors reduce both the spectral-placement and information losses. All three remain below the sinusoidal benchmark, which is obtained by the separate coherence limit of Sec.~\ref{sec:maxsignal}.

\paragraph{Power-law improvement and fixed-rate reversals.}
Figure~\ref{fig:valueheatmap} shows the dependence on both parameters: moving right increases signal speed, and moving down increases smoothness. Color gives value relative to each market's bound; labelled contours give the percentage of the same signal's anticipative value lost to causality.

In the power-law panel~(b), the dashed curve follows the maximizing decay rate. Moving toward smoother predictors along this curve raises value and, for the displayed parameters, raises the preferred rate. The orange point is the OU optimum. With $\nu_m=3/4$ and $\beta=\gamma=s_\alpha^2=1$, the predictor $\nu_s=-2$ has preferred rate $\theta_s\simeq1.4064$ and value $0.2122$, compared with $1/3$ and $0.1624$ for OU. The full Mat\'ern supremum is approximately $0.2314$, approached in the Gaussian spectral limit described in Appendix~\ref{app:valuedetails}.

A vertical slice holds speed fixed and can give the opposite ordering. At $\theta_s=0.01$, values for $\nu_s=-0.4,0,0.3$ are approximately $0.0365,0.0490,0.0730$: the rougher signal is more valuable because less variance lies at very low frequencies, where replenishment is costly. This is why the general smoothness comparison optimizes the rate separately for each exponent.

The contours in panel~(b) show that fast, rough signals lose a substantial fraction of their anticipative value. Slow, smooth signals have a smaller information gap but also low value, because their spectral placement is costly even with full foresight.

\paragraph{Equality under exponential impact.}\label{sec:informationgap}
For pure exponential impact, $\widehat g=\gamma/(\theta_m^2+\omega^2)$, take $\theta_m,\beta>0$ and $\lambda=\eta=0$. In the completed strategy space, every Mat\'ern exponent has the same rate-optimized supremum:
\begin{equation}\label{eq:expoptimizedvalue}
\bar v^\star(\nu_s)=\frac{s_\alpha^2}{2\gamma}
\max\left\{1,\frac{\theta_m^2}{\beta^2}\right\}.
\end{equation}
The supremum is approached as $\theta_s\to\infty$ for $\theta_m<\beta$ and as $\theta_s\to0$ for $\theta_m>\beta$. At $\theta_m=\beta$, every rate attains it with $\pi^\star=\alpha/\gamma$; for OU input this has no ordinary finite-variance trading rate, but causal smoothing approaches its value. In panel~(a), $\theta_m<\beta$: faster signals approach the high-frequency friction minimum with vanishing causality loss. Slow, smooth signals also have a small information gap, but occupy costlier frequencies.

\section{Scope and extensions}\label{sec:extensions}

\paragraph{Finite trading horizons.}\label{sec:liquidation}\label{sec:initialhistory}
On a finite trading window, we specify initial inventory, the impact left by realized trades, and terminal conditions. Inherited impact enters as a known contribution to future marginal trading cost, alongside local execution costs and inventory risk. Under suitable coercivity and regularity conditions, causal triangular factorization gives the corresponding constrained projection \citep{GohbergKrein1970}. In Appendix~\ref{sec:boundaries} we summarize this formulation and its relation to exponential-impact endpoint blocks \citep{ObizhaevaWang2013}, power-law liquidation profiles \citep{GatheralSchiedSlynko2012}, and Gaussian signal-adaptive fractional execution \citep{FordeSanchezSmith2022}. We also give an explicit half-line continuation from realized trades for the pure-power-law OU case in Appendix~\ref{app:initialhistory}. General Volterra execution and cross-impact portfolio choice are treated in \citet{AbiJaberNeuman2025,AbiJaberNeumanTuschmann2026}.

\paragraph{Causally separable convex costs.}\label{sec:nonlinear}
For a convex friction functional, optimal value is a constrained convex conjugate \citep{Rockafellar1970}. The explicit forecast-and-response construction extends when a causal linear isomorphism $B$ separates the cost as $\Phi(\pi)=\E[\phi((B\pi)_t)]$. We then maximize pointwise against the conditional forecast and apply $B^{-1}$. In Appendix~\ref{app:nonlinear} we state the range, integrability, and attainment conditions. The isomorphism must respect the chosen stationary space; an unrestricted-mean nonlinear problem cannot automatically use the power-law factor, whose inverse annihilates the zero-frequency component. General nonlinear transient costs require establishing a suitable representation separately.

\section{Conclusion}\label{sec:outlook}

We have obtained a causal stationary trading rule by using Wiener--Hopf factorization to separate forecasting from causal execution. For OU signals, the forecast is proportional to the current observation, so we can examine the market response directly.

We obtain an explicit fractional trading policy under power-law impact even for this one-state predictor. Its nonrational transfer precludes an exact finite-dimensional linear time-invariant realization. Exponential impact admits finite-dimensional execution dynamics, while history-dependent Mat\'ern forecasts can make the full trading response nonrational. Predictor memory changes the response even with local execution costs. We treat both sources of memory with the same factorization.

Exposure bleeding makes maintaining a position costly. In the power-law examples, slow signals require sustained replenishment, while fast signals require costly adjustments. We therefore find an intermediate OU signal speed that gives the highest net value. For ordinary holdings with an inventory penalty, slower signals are favored in the markets we study.

We have used stationary trading to study the value of information in systems with memory. Even for fully predictable signals, value depends on the relation between their temporal structure and the dynamics of the response. Extending the analysis beyond stationary quadratic models would clarify which aspects of this dependence persist under more general dynamics.

\begin{acknowledgments} The author thanks CFM colleagues, and in particular J.-Y. Audibert, A. Bambade, R. Bénichou, J.-P. Bouchaud, R. Maillet, C. Masselin, I. Mastromatteo and M. Potters, for useful discussions. Research, literature review, and manuscript preparation were assisted by LLMs.
\end{acknowledgments}

\appendix
\section{Derivation of the trading filters}\label{app:filterderivations}
\subsection{Domain and projection}\label{app:outerfactor}
In this appendix we justify the forecast-and-response construction of Section~II and evaluate it for the models in Section~III. We first specify the admissible positions: finite quadratic cost can allow rougher positions than a finite-variance trading rate would. This distinction matters in the pure-impact limits.

We use a real Hilbert space $\mathcal H$ of square-integrable stationary observables, invariant under a strongly continuous, measure-preserving shift group. The filtration is shift-covariant, conditional expectation preserves $\mathcal H$, and $\mathcal M=\mathcal H\cap L^2(\F_0)$ is the adapted subspace. For the spectral measure $E_x$, normalized by $E_x(\R)=\|x\|_2^2$, define
\begin{equation}\label{eq:frictiondomain}
\|x\|_F^2=\tfrac12\int_\R\widehat f\,dE_x,
\qquad \mathcal A=\{x\in\mathcal M:\|x\|_F<\infty\}.
\end{equation}
Assume $m\leq\widehat f<\infty$ almost everywhere for every $E_x$, with $m>0$, and choose conjugate-symmetric boundary versions of its outer factor satisfying $|\widehat f_+|^2=\widehat f$. The closed multiplier $\mathsf F_+$ has bounded inverse $\mathsf F_+^{-1}$, whose adjoint is $\mathsf F_-^{-1}$. The logarithmic condition \eqref{eq:szego} gives Lebesgue-almost-everywhere factorization \citep{Bingham2021}; compatibility at singular spectral components is an additional assumption.

\begin{proposition}[Adapted optimization]\label{prop:formaloptimization}
If $\mathcal A$ is dense in $\mathcal M$ and $\mathsf F_+\mathcal A=\mathcal M$, then \eqref{eq:solution}--\eqref{eq:gap} hold, with the compression interpreted through the closed form \eqref{eq:frictiondomain}.
\end{proposition}
\begin{proof}
Writing $\xi=\mathsf F_-^{-1}\alpha$ and $\pi=\mathsf F_+^{-1}y$ gives, for $y\in\mathcal M$,
\[
J(\mathsf F_+^{-1}y;\alpha)=\tfrac12\|\mathsf P\xi\|_2^2
-\tfrac12\|y-\mathsf P\xi\|_2^2.
\]
Thus $\pi^\star=\mathsf F_+^{-1}\mathsf P\mathsf F_-^{-1}\alpha$. Its weak normal equation is $\langle\mathsf F_+\pi^\star,\mathsf F_+h\rangle=\langle\alpha,h\rangle$ for $h\in\mathcal A$; taking $h=\pi^\star$ gives the gain identity. Allowing $y\in\mathcal H$ gives the anticipative value, and orthogonality gives the gap.
\end{proof}

To apply the proposition to our examples, we approximate admissible positions by causally smoothed positions. This also relates the completed strategy space to policies with ordinary trading rates.

For the explicit markets with $\widehat f\leq C(1+\omega^2)$, the causal regularizer $\widehat R_\varepsilon=(1-i\varepsilon\omega)^{-1}$ verifies the hypotheses: $\mathsf F_+R_\varepsilon$ and $\mathsf F_+^{-1}$ are bounded causal multipliers, and $R_\varepsilon\to I$ strongly and in form norm. These facts give density and both inclusions in $\mathsf F_+\mathcal A=\mathcal M$ \citep{Staffans1995}. For power-law impact with $\beta>0$, $\widehat f(0)=\infty$: restrict to $E_x(\{0\})=0$, assume $\mathsf P$ preserves this space, and use
\[
\widehat Q_\varepsilon(\omega)=
\frac{-i\omega}{(\varepsilon-i\omega)(1-i\varepsilon\omega)}
\]
in the same argument. The centered Gaussian examples satisfy this restriction. For $\beta=0$, $\lambda>0$, use the continuous extension $\widehat f(0)=\lambda$. In all cases $R_\varepsilon$ approximates finite-form positions by positions with finite-variance rates. Thus $\mathcal A$ is their friction-norm completion. The execution mass $\eta\delta_0$ has half-weight in the causal cost integral; masses in forecast and reconstruction convolutions have full weight.

For Gaussian predictors under the information assumption of Sec.~\ref{sec:model}, independence of future innovations makes conditional expectation retain the nonnegative-lag kernel of $S_{\alpha,+}/\widehat f_-$. Applying $\mathsf F_+^{-1}$ proves \eqref{eq:signaltransfer}; Parseval gives \eqref{eq:innovations}.

\subsection{Bessel and cutoff formulas}\label{app:besselmarkets}
We now evaluate the causal projection for predictor and impact spectra in the same Bessel family. Working with both exponents at once lets us recover the OU and exponential special cases and identify how the two sources of memory combine.

Take $\beta,\gamma,\theta_s>0$, $\lambda=0$, $\nu_s<1/2$, and $0\leq\nu_m<1$ for $\theta_m>0$, or $1/2<\nu_m<1$ for $\theta_m=0$. The spectra and factors are
\begin{equation}\label{eq:besselnormalization}
\begin{gathered}
\widehat g=\gamma(\theta_m^2+\omega^2)^{\nu_m-1},\qquad
S_{\alpha,+}=a(\theta_s-i\omega)^{\nu_s-1},\\
a^2=\frac{2\sqrt\pi\,\Gamma(1-\nu_s)}{\Gamma(1/2-\nu_s)}
s_\alpha^2\theta_s^{1-2\nu_s},\\
\widehat f_+(z)=\sqrt\gamma\frac{\beta-iz}{(\theta_m-iz)^{1-\nu_m}}.
\end{gathered}
\end{equation}
Powers are positive on the positive imaginary axis. The causal kernel of $(\theta-i\omega)^{\nu-1}$ is $t^{-\nu}e^{-\theta t}/\Gamma(1-\nu)$ for $t>0$ \citep{AronszajnSmith1961}. To compute the forecast kernel, we integrate the predictor's innovations kernel against the fractional part of the inverse friction factor. Let $h_\alpha$ have Fourier transform $S_{\alpha,+}$ and, for $\nu_m>0$, define
\[
\varphi(t)=\frac1{\Gamma(\nu_m)}\int_0^\infty
r^{\nu_m-1}e^{-\theta_m r}h_\alpha(t+r)\,dr.
\]
Euler's integral \citep[Eq.~15.6.1]{DLMF} gives
\begin{equation}\label{eq:besselprojectionaux}
\begin{gathered}
\widehat\varphi(\omega)=C\,{}_2F_1\!\left(
\begin{matrix}1,1+\nu_m-\nu_s\\1+\nu_m\end{matrix};
\frac{\theta_m+i\omega}{\theta_s+\theta_m}\right),\\
C=\frac{a\,\Gamma(1+\nu_m-\nu_s)}
{\Gamma(1-\nu_s)\Gamma(1+\nu_m)(\theta_s+\theta_m)^{1+\nu_m-\nu_s}}.
\end{gathered}
\end{equation}
At $\nu_m=0$, use $\widehat\varphi=S_{\alpha,+}$; analytic continuation from positive imaginary frequencies fixes the branch. Decomposing $1/\widehat f_+$ into $(\theta_m-i\omega)^{-\nu_m}[1+(\theta_m-\beta)/(\beta-i\omega)]/\sqrt\gamma$ and projecting gives
\begin{equation}\label{eq:besseljointtransfer}
\begin{aligned}
\widehat k(\omega)={}&\frac{(\theta_m-i\omega)^{1-\nu_m}}
{\gamma(\beta^2+\omega^2)S_{\alpha,+}(\omega)}\\
&\times\bigl[(\theta_m+i\omega)\widehat\varphi(\omega)
-(\theta_m-\beta)\widehat\varphi(i\beta)\bigr].
\end{aligned}
\end{equation}
The apparent pole at $i\beta$ is removable. For OU input, $h_\alpha(t+r)=e^{-\theta_s r}h_\alpha(t)$, so $\widehat q=S_{\alpha,+}/\widehat f_+(i\theta_s)$, where $\widehat q=[S_{\alpha,+}/\widehat f_-]_+$. This proves the OU and fractional formulas in Sec.~III. For local friction, $\widehat q=[S_{\alpha,+}(\omega)-S_{\alpha,+}(ib)]/[\sqrt\eta(b+i\omega)]$, proving \eqref{eq:maternlocaltransfer}.

For Fig.~\ref{fig:filters}, we retain instantaneous execution costs so that we can follow the crossover from fractional response to the high-frequency cutoff. We choose a regularization with an explicit spectral factor: take $\eta>0$, $\Omega>\theta_m$, and
\begin{equation}\label{eq:cutoffexecution}
\widehat g(\omega)=\eta\left(\frac{\Omega^2+\omega^2}
{\theta_m^2+\omega^2}\right)^{1-\nu_m},
\quad \gamma=\eta\Omega^{2(1-\nu_m)}.
\end{equation}
\begin{equation}\label{eq:cutofffactor}
\widehat f_+(z)=\sqrt\eta(\beta-iz)
\left(\frac{\Omega-iz}{\theta_m-iz}\right)^{1-\nu_m}.
\end{equation}
The pure-impact limit fixes $\gamma$ and sends $\Omega\to\infty$. With $A(r)=\widehat f_+(ir)$ and $H_\omega(r)=[A(r)(r-i\omega)]^{-1}$, choose an integer $n\geq0$ with $n+\nu_s>0$. The gamma integral for $h_\alpha$ gives
\begin{equation}\label{eq:maternquadrature}
\begin{aligned}
\widehat k(\omega)={}&\frac{(-1)^n(\theta_s-i\omega)^{1-\nu_s}}
{\widehat f_+(\omega)\Gamma(1-\nu_s)\Gamma(n+\nu_s)}\\
&\times\int_{\theta_s}^\infty(r-\theta_s)^{n+\nu_s-1}
H_\omega^{(n)}(r)\,dr.
\end{aligned}
\end{equation}
This representation covers every $\nu_s<1/2$; $H_\omega^{(n)}=O(r^{-n-2})$ ensures convergence at infinity. For $q(0+)\ne0$, $\widehat q\sim q(0+)/(-i\omega)$ gives \eqref{eq:maternplscaling} and the cutoff tail $\widehat k\sim q(0+)(-i\omega)^{-1-\nu_s}/(a\sqrt\eta)$. The plotted coefficients are positive; cancellation requires the next asymptotic term.

\section{Signal values and comparisons}\label{app:valuedetails}
\subsection{Value and rate optimization}
In this appendix we derive the comparisons in Section~\ref{sec:value}. Once we have the projected forecast kernel from Appendix~\ref{app:filterderivations}, its squared norm gives net value. We can therefore compare signals without simulating their realized trading profits.

Under the assumptions of Appendix~\ref{app:besselmarkets}, the innovations isometry gives
\begin{equation}\label{eq:pureimpactvalue}
v^\star=\tfrac12\int_0^\infty q(t)^2dt
=\frac1{2\pi}\int_0^\infty|\widehat q(\omega)|^2d\omega.
\end{equation}
This is a one-dimensional quadrature; rate optimization includes both endpoint limits. For OU input,
\begin{equation}\label{eq:besselsolution}
v^\star(0,\theta_s)=\frac{s_\alpha^2}{2\gamma}
\frac{(\theta_s+\theta_m)^{2-2\nu_m}}{(\theta_s+\beta)^2}.
\end{equation}
For $\nu_m>0$, logarithmic differentiation gives the maximizing rate $[(1-\nu_m)\beta-\theta_m]/\nu_m$ when positive, and the zero-rate supremum otherwise. At $\nu_m=0$, value is monotone with sign $\beta-\theta_m$. The cutoff multiplies \eqref{eq:besselsolution} by $(1+\theta_s/\Omega)^{-2(1-\nu_m)}$; for $\theta_m=0$, the optimal rate solves $\theta_s^2+\nu_m\Omega\theta_s=(1-\nu_m)\beta\Omega$.

For pure exponential impact, we also compute the information gap to establish when causal trading approaches the anticipative bound. Put $r=\theta_s/\beta$ and $r_m=\theta_m/\beta$. The negative-lag innovations kernel is $a(\theta_m-\beta)e^{\beta t}/[\sqrt\gamma(\theta_s+\beta)^{1-\nu_s}]$. Half its squared norm gives
\begin{equation}\label{eq:expinformationgap}
v_{\rm ant}-v^\star=\frac{s_\alpha^2}{2\gamma}
\frac{\sqrt\pi\Gamma(1-\nu_s)}{\Gamma(1/2-\nu_s)}
\frac{(r_m-1)^2r^{1-2\nu_s}}{(1+r)^{2-2\nu_s}}.
\end{equation}
The gap vanishes at both rate limits. Since $1/\widehat f=\gamma^{-1}[1+(\theta_m^2-\beta^2)/(\beta^2+\omega^2)]$, dominated convergence gives the corresponding anticipative limits $s_\alpha^2r_m^2/(2\gamma)$ and $s_\alpha^2/(2\gamma)$. The bound $1/\widehat f\leq\gamma^{-1}\max\{1,r_m^2\}$ then proves \eqref{eq:expoptimizedvalue}.

\subsection{Smoothness and inventory risk}
To compare smoothness after optimizing speed, we express a rougher predictor spectrum as a mixture of smoother spectra at different rates. A fixed trading filter has an objective linear in the spectrum, which lets us compare the optimized values through this mixture.

Write $\rho=1-\nu_s$ and normalize $S_\alpha=2\pi s_\alpha^2p^{(\theta,\rho)}$, where
\[
p^{(\theta,\rho)}(\omega)=\frac{\Gamma(\rho)\theta^{2\rho-1}}
{\sqrt\pi\Gamma(\rho-1/2)(\theta^2+\omega^2)^\rho}.
\]
For $\rho_2>\rho_1>1/2$, Euler's beta integral \citep[Eq.~15.6.1]{DLMF} gives
\begin{equation}\label{eq:besselscalemixture}
\begin{gathered}
p^{(\theta,\rho_1)}=\E[p^{(\theta/\sqrt U,\rho_2)}],\\
U\sim\operatorname{Beta}(\rho_1-\tfrac12,\rho_2-\rho_1).
\end{gathered}
\end{equation}
For each causal response $\ell$, its objective $\mathcal J_\ell(S)=(2\pi)^{-1}\int(\operatorname{Re}\widehat\ell-\widehat f|\widehat\ell|^2/2)S$ is linear in $S$. Gaussian linear optimality therefore implies
\begin{equation}\label{eq:exponentmixturebound}
\begin{aligned}
v^\star(1-\rho_1,\theta)&\leq\E[v^\star(1-\rho_2,\theta/\sqrt U)]\\
&\leq\sup_{\theta'>0}v^\star(1-\rho_2,\theta').
\end{aligned}
\end{equation}
A finite-friction filter under the mixture is admissible for almost every component; approximation extends the inequality to the completed space. Taking the supremum over $\theta$ proves the smoothness ordering and \eqref{eq:preferredexponent}.

We identify the full Mat\'ern supremum through its Gaussian spectral limit:
\begin{equation}\label{eq:materngaussianlimit}
\sup_{\nu_s<1/2,\,\theta_s>0}v^\star
=\sup_{\sigma>0}\frac{s_\alpha^2}{2\sqrt\pi\sigma}
\int_\R\frac{e^{-\omega^2/\sigma^2}}{\widehat f(\omega)}\,d\omega.
\end{equation}
Indeed, $p^{(\theta,\rho)}$ mixes Gaussian densities $e^{-\omega^2/\sigma^2}/(\sqrt\pi\sigma)$ with $\sigma=\theta/\sqrt Z$, $Z\sim\operatorname{Gamma}(\rho-1/2,1)$, proving the upper bound. Conversely, fix $\sigma$, set $\theta=\sigma\sqrt\rho$ and $d_\rho=\sqrt\rho/\sigma$. The centered factors $G_\rho(\omega)=e^{-i\omega d_\rho}S_{\alpha,+}(\omega)$ converge in $L^2$ to $(2\sqrt\pi s_\alpha^2/\sigma)^{1/2}e^{-\omega^2/(2\sigma^2)}$. The inverse transforms $\psi_\rho$ of $G_\rho/\widehat f_-$ converge to $\psi$ in $L^2$ by boundedness. The uncentered negative-lag energy $\int_{-\infty}^{-d_\rho}|\psi_\rho(t)|^2dt\to0$, proving adapted attainability. The limiting Gaussian spectrum itself need not admit an outer innovations factor.

We finally examine ordinary holdings with inventory risk, to distinguish the smoothness comparison from the effect of replenishing bleeding exposure. For $\beta=0$, $\lambda>0$, the execution models considered here have $\widehat f=\lambda+\omega^2\widehat g$ nondecreasing in $|\omega|$. Rescaling a causal response as $L(\omega/\theta_s)$ leaves signal gain unchanged and replaces its friction by $\widehat f(\theta_s\omega)$, so value decreases with $\theta_s$. Since $\theta_s/\sqrt U\geq\theta_s$, \eqref{eq:exponentmixturebound} also orders smoothness at fixed rate. As $\theta_s\downarrow0$, the trial position $\operatorname{EMA}_b[\alpha]/\lambda$, $b>0$, has vanishing execution cost and value tending to $s_\alpha^2/(2\lambda)$. The execution integral converges because $\omega^2\widehat g/(b^2+\omega^2)$ is bounded and vanishes at zero. Together with \eqref{eq:signalbound}, this proves the stated supremum for every exponent. With $\lambda=0$ as well, sinusoidal values diverge as their frequency tends to zero.

\subsection{Sharpness of the signal bound}
We establish both attainment by a perfectly predictable signal and approximation by signals with ongoing innovations. The latter shows how increasing coherence can bring attainable value close to the bound.

The bound and sinusoidal attainment follow directly from \eqref{eq:signalbound}--\eqref{eq:maxsignalpolicy}, using spectral measures for line spectra. Differentiating $\widehat f$ gives \eqref{eq:maxsignalcomparison}; with cutoff power-law impact, the minimizing frequency solves $\omega_*^4+\nu_m\Omega^2\omega_*^2=(1-\nu_m)\beta^2\Omega^2$. Absolute continuity makes the bound strict for square-integrable innovations kernels in these markets. To approach it, use damped Gaussian oscillators with spectra
\begin{equation}\label{eq:maxsignalapprox}
S_\delta(\omega)=\frac{4\delta\omega_*^2s_\alpha^2}
{(\omega_*^2-\omega^2)^2+4\delta^2\omega^2},\qquad\delta>0.
\end{equation}
Their variance is $s_\alpha^2$ and their spectra concentrate at $\pm\omega_*$. For $\mathcal E_\delta=(2\pi)^{-1}\int\widehat fS_\delta$, the adapted trial $\pi=s_\alpha^2\alpha/\mathcal E_\delta$ has value $s_\alpha^4/(2\mathcal E_\delta)$. Away from resonance, the integrand is bounded by constants times $\delta|\omega|^{2\nu_m-2}$ near zero and $\delta|\omega|^{-2}$ at infinity, including positive cutoff cost. Hence $\mathcal E_\delta\to s_\alpha^2\widehat f(\omega_*)$, proving sharpness within the innovations class.

\section{Initial conditions, finite windows, and liquidation}\label{sec:boundaries}\label{app:liquidation}
\subsection{Finite windows}
The stationary solution presumes an established trading history. In this appendix we consider a trader starting from a specified inventory and inherited impact, with a terminal requirement when needed. We first formulate the finite-window problem, then derive an explicit power-law continuation on the half-line.

On $[0,T]$, write $\pi_t=e^{-\beta t}\pi_0+(L_\beta u)(t)$, where $L_\beta u=\int_0^t e^{-\beta(t-s)}u_sds$, and let $h_{\rm past}(t)=\int_{-\infty}^0g(t-s)u_s^{\rm past}ds$. Assume square-integrable data, with $\pi_0,h_{\rm past}$ known at time zero. Up to terms independent of $u$, the objective is $\E\langle b_T,u\rangle-\E\langle u,\mathsf F_Tu\rangle/2$, with
\begin{equation}\label{eq:windowfriction}
\begin{gathered}
b_T=L_\beta^*\alpha-\lambda L_\beta^*(e^{-\beta\cdot}\pi_0)-h_{\rm past},\\
\mathsf F_T=G_T+\lambda L_\beta^*L_\beta,\qquad
(G_Tu)(t)=\int_0^Tg(|t-s|)u_sds.
\end{gathered}
\end{equation}
Here $G_T$ includes $\eta I$. With $\eta>0$ and a continuous positive-semidefinite regular kernel, triangular factorization gives $\mathsf F_T=\mathsf F_{T,+}^*\mathsf F_{T,+}\geq\eta I$ with causal factor and inverse \citep{GohbergKrein1970}. Completing the square gives
\begin{equation}\label{eq:finiteprojection}
u^\star=\mathsf F_{T,+}^{-1}
\operatorname{Proj}_{\mathsf F_{T,+}\mathcal U}
\bigl((\mathsf F_{T,+}^*)^{-1}b_T\bigr),
\end{equation}
where $\mathcal U$ is the nonempty closed affine set of adapted rates satisfying the prescribed terminal condition, if any. A target $\pi_T$ imposes $\int_0^T e^{-\beta(T-t)}u_tdt=\pi_T-e^{-\beta T}\pi_0$. Without this constraint, the projection is $\mathsf P$, and conditioning the normal equation on $\F_0$ gives
\[
\E[u^\star\mid\F_0]=\mathsf F_T^{-1}\E[b_T\mid\F_0].
\]
An $\F_0$-measurable $L^2$ right trace determines the initial rate. Pure impact ($\eta=0$) requires the corresponding finite-energy domain.

As boundary-condition benchmarks, we recall the signal-free liquidation profiles. They show how exponential and power-law impact lead to different behavior near the trading endpoints. For liquidation with $\beta=0$, $\pi_0=x$, $\pi_T=0$, zero signal, no inherited impact, and $\lambda=\eta=0$, the classical profiles are \citep{ObizhaevaWang2013,GatheralSchiedSlynko2012}
\begin{equation}\label{eq:liquidationprofiles}
\begin{aligned}
U_{\rm exp}(dt)&=-\frac{x[\delta_0(dt)+\theta_mdt+\delta_T(dt)]}{2+\theta_m T},\\
u_{\rm pow}^\star(t)&=-\frac{x[t(T-t)]^{\nu_m-1}}
{T^{2\nu_m-1}\mathrm B(\nu_m,\nu_m)}.
\end{aligned}
\end{equation}
Here $\mathrm B$ is the beta function and $1/2<\nu_m<1$ in the power-law case. Gaussian signal-adaptive fractional liquidation is treated by \citet[Theorem~2.2]{FordeSanchezSmith2022}.

\subsection{Power-law continuation}\label{app:initialhistory}
We now retain the OU predictor and ask how realized past trades change the fractional policy after time zero. We separate the response to the signal from the correction for inherited impact, then compute the initial trading behavior from their mismatch.

Take $\widehat g=\gamma|\omega|^{2\nu_m-2}$, $1/2<\nu_m<1$, $\lambda=\eta=0$, $\beta,\gamma,\theta_s>0$, and $d\alpha_t=-\theta_s\alpha_tdt+a\,dW_t$ on $[0,\infty)$, with $a$ as in \eqref{eq:besselnormalization}. The initial $\alpha_0,\pi_0$ are square-integrable and $\F_0$-measurable; future Brownian increments are independent of $\F_0$. Bounded, $\F_0$-measurable past rates of finite duration suffice for the formulas below. Set
\begin{equation}\label{eq:initialmemory}
\begin{aligned}
c_s&=\frac{\theta_s^{1-\nu_m}}{\gamma(\theta_s+\beta)},\\
\phi_{\rm past}(t)&=\frac1{\Gamma(1-\nu_m)}
\int_{-\infty}^0(t-s)^{-\nu_m}u_s^{\rm past}ds.
\end{aligned}
\end{equation}
The optimal continuation is
\begin{equation}\label{eq:initializedfractional}
\begin{aligned}
u_t^\star={}&c_sD_{0+}^{1-\nu_m}\alpha_t\\
&-\frac{\sin(\pi\nu_m)}{\pi}t^{\nu_m-1}
\int_{-\infty}^0\frac{(-s)^{1-\nu_m}}{t-s}u_s^{\rm past}ds,
\end{aligned}
\end{equation}
where $D_{0+}^{1-\nu_m}f=\Gamma(\nu_m)^{-1}\partial_t\int_0^t(t-s)^{\nu_m-1}f(s)ds$ is the initialized Riemann--Liouville derivative \citep{SamkoKilbasMarichev1993}. Initial inventory contributes fixed earnings $\alpha_0\pi_0/(\theta_s+\beta)$ and leaves the rate unchanged. For $\Delta_0=c_s\alpha_0-\phi_{\rm past}(0)$,
\begin{equation}\label{eq:initialtrade}
\begin{aligned}
u_t^\star&=\frac{\Delta_0}{\Gamma(\nu_m)}t^{\nu_m-1}
+o_{\mathbb P}(t^{\nu_m-1}),\\
\int_0^\varepsilon u_t^\star dt
&=\frac{\Delta_0}{\Gamma(1+\nu_m)}\varepsilon^{\nu_m}
+o_{\mathbb P}(\varepsilon^{\nu_m}).
\end{aligned}
\end{equation}
Thus the mismatch determines initial buying or selling. A nonzero block has infinite impact energy, so $\pi_{0+}=\pi_0$. If $\Delta_0=0$ and the past rate has a left limit, $u_t^\star\to u_{0-}^{\rm past}$ in probability.

For verification, beta integration factors the kernel as
\[
g(|t-s|)=\frac{\gamma}{\Gamma(1-\nu_m)^2}
\int_{\max(t,s)}^\infty(r-t)^{-\nu_m}(r-s)^{-\nu_m}dr.
\]
Write $\E_t=\E[\,\cdot\mid\F_t]$ and $y=\phi_{\rm past}+I_{0+}^{1-\nu_m}u$. The adapted marginal cost is $\gamma\E_t\int_t^\infty(r-t)^{-\nu_m}y_rdr/\Gamma(1-\nu_m)$. At $y=c_s\alpha$, it equals the marginal earnings $\alpha_t/(\theta_s+\beta)$. Applying $D_{0+}^{1-\nu_m}$ gives \eqref{eq:initializedfractional}. Here $I_{0+}^{1-\nu_m}u=\Gamma(1-\nu_m)^{-1}\int_0^t(t-s)^{-\nu_m}u_sds$.

Optimality is relative: let $J_T$ be conditional expected earnings minus causal impact costs on $[0,T]$, including inherited impact. For bounded adapted perturbations $w$ of finite duration, the first variation vanishes and
\begin{equation}\label{eq:halflinerelative}
\begin{gathered}
\lim_{T\to\infty}[J_T(u^\star+w)-J_T(u^\star)]\\
=-\frac\gamma2\E_0\int_0^\infty
|I_{0+}^{1-\nu_m}w|^2dt.
\end{gathered}
\end{equation}
The identity gives uniqueness in this comparison class and extends to finite-energy perturbations with convergent cross terms. This criterion permits nonstationary initial data while the undiscounted total OU reward generally diverges.

\section{Nonlinear costs}\label{app:nonlinear}
For causally separable convex costs, we forecast, optimize pointwise in transformed positions, and reconstruct the policy.

Let $X$ be a stationary space with closed adapted subspace $\mathcal M$, and assume $\alpha\in X^*$ under the payoff pairing $\langle\alpha,\pi\rangle=\E[\alpha_t\pi_t]$. For proper lower-semicontinuous convex $\Phi$, value is $(\Phi+\iota_{\mathcal M})^*(\alpha)$, with $\iota_{\mathcal M}$ the indicator and $*$ convex conjugacy \citep{Rockafellar1970}. Suppose $\Phi(\pi)=\E[\phi((B\pi)_t)]$ for proper lower-semicontinuous convex $\phi$. Assume $B:X\to L^2$ is a bounded shift-commuting isomorphism, $B,B^{-1}$ are causal, and $B\mathcal M$ is the adapted stationary $L^2$ subspace. Set
\begin{equation}\label{eq:convexresponse}
\begin{gathered}
\xi=(B^{-1})^*\alpha,\qquad \zeta=\mathsf P\xi,\\
y_t^\star\in\partial\phi^*(\zeta_t),\qquad
\pi^\star=B^{-1}y^\star,\\
v^\star=\E[\phi^*(\zeta_t)].
\end{gathered}
\end{equation}
Here $\partial\phi^*$ is the convex subdifferential. The formula holds whenever a measurable selection $y^\star\in B\mathcal M$ exists with $\E|\phi(y_t^\star)|<\infty$; strict convexity of $\phi$ gives uniqueness. In $y=B\pi$, the objective is $\E[\zeta_ty_t-\phi(y_t)]$, so pointwise conjugacy proves optimality. Conditional Jensen compares this value with $\E[\phi^*(\xi_t)]$ when finite. The space must admit the resulting means. For power-law friction with $\beta>0$, $\mathsf F_+^{-1}$ annihilates constants and fails this unrestricted-mean isomorphism condition. General nonlinear transient costs require a separately established separable representation.

\makeatletter
\patchcmd{\bibsection}{\onecolumngrid@push}
  {\onecolumngrid@push\vspace{10pt}}{}{\errmessage{Bibliography spacing patch failed}}
\apptocmd{\thebibliography}{\interlinepenalty=10000}{}
  {\errmessage{Bibliography line-break patch failed}}
\makeatother
\bibliography{references}
\end{document}